\documentclass{article}
\usepackage{fullpage}
\usepackage{graphicx}
\usepackage{float}
\usepackage{color}
\usepackage[skip=2pt,font=footnotesize]{caption}
\usepackage{listings}
\usepackage{authblk}
\usepackage{placeins}

\newcommand{\csp}{{~}} 
\usepackage{multicol}

\begin{document}

\begingroup
\centering
{\LARGE Tracking Human Mobility using WiFi signals} \\[1.5em]

\begin{table}[h!]
\centering
\begin{tabular}{c c c c}
Piotr Sapiezynski*$^{1}$ & Arkadiusz Stopczynski$^{1,2}$ & Radu Gatej$^{3}$ & Sune Lehmann$^{1,4}$ \\[0em]
pisa@dtu.dk & arks@mit.edu & radu.gatej@econ.ku.dk & sljo@dtu.dk
\end{tabular}
\end{table}
\endgroup

\noindent $^{1}$ Department of Applied Mathematics and Computer Science, Technical University of Denmark\\
$^{2}$ Media Lab, Massachusetts Institute of Technology \\
$^{3}$ Department of Economics, University of Copenhagen \\
$^{4}$ Niels Bohr Institute, University of Copenhagen \\

\section*{Abstract}
We study six months of human mobility data, including WiFi and GPS traces recorded with high temporal resolution, and find that time series of WiFi scans contain a strong latent location signal.
In fact, due to inherent stability and low entropy of human mobility, it is possible to assign location to WiFi access points based on a very small number of GPS samples and then use these access points as location beacons.
Using just one GPS observation per day per person allows us to estimate the location of, and subsequently use, WiFi access points to account for 80\% of mobility across a population.
These results reveal a great opportunity for using ubiquitous WiFi routers for high-resolution outdoor positioning, but also significant privacy implications of such side-channel location tracking.

\section*{Introduction}

Due to the ubiquity of mobile devices, the collection of large-scale, longitudinal data about human mobility is now commonplace\csp\cite{lazer2009life}.
High-resolution mobility of individuals and entire social systems can be captured through a multitude of sensors available on modern smartphones, including GPS and sensing of nearby WiFi APs (access points or routers) and cell towers. 
Similarly, mobility data may be collected from systems designed to enable communication and connectivity, such as mobile phone networks or WiFi systems (e.g. at airports or on company campuses)\csp\cite{lim2007real,ferris2006gaussian}.
Additionally, large companies such as Google, Apple, Microsoft, or Skyhook, combine WiFi access points with GPS data to improve positioning\csp\cite{skyhook}, a practice known as `wardriving'.
While widely used, the exact utility and mechanics of wardriving are largely unknown, with only narrow and non-systematic studies reported in the literature\csp\cite{rekimoto2007lifetag, kawaguchi2009wifi}.
As a consequence, it is generally not known how WiFi networks can be used for sensing mobility on a societal scale; this knowledge is proprietary to large companies.


In the scientific realm, the mobility patterns of entire social systems are important for modeling spreading of epidemics on multiple scales: metropolitan networks\csp\cite{eubank2004modelling, sun2014efficient,liang2013unraveling} and global air traffic networks\csp\cite{colizza2007, hufnagel2004forecast}; traffic forecasting\csp\cite{kitamura2000micro}; understanding fundamental laws governing our lives, such as regularity\csp\cite{gonzalez2008understanding}, stability\csp\cite{Lu18062012}, and predictability\csp\cite{song2010limits}.
Predictability and stability of human mobility are also exploited by commercial applications such as intelligent assistants; for example Google Now\csp\cite{google_now} is a mobile application, which learns users' habits to, among other services, conveniently provide directions to the next inferred location.


Mobility traces are highly unique and identify individuals with high accuracy\csp\cite{de2013unique}.
Sensitive features can be extracted from mobility data, including home and work locations, visited places, or personality traits\csp\cite{de2013predicting}. 
Moreover, location data are considered the most sensitive of all the commonly discussed personal data collected from or via mobile phones\csp\cite{staiano2014money}.


Here, we show that a time sequence of WiFi access points is effectively equal to location data.
Specifically, having collected both GPS and WiFi data with high temporal resolution (median of 5 minutes for GPS and 16 seconds for WiFi) in a large study\csp\cite{10.1371/journal.pone.0095978}, we use six months of data for 63 participants to model how lowering the rate of location sampling influences our ability to infer mobility.
The study participants are students with heterogeneous mobility patterns.
They all attend lectures on campus located outside of the city center, but live in dormitories and apartments scattered across the metro area at various distances from the university.



By mapping the WiFi data, we are able to quantify details of WiFi-based location tracking, which are usually not available to the general public.
We find that the geo-positioning inferred from WiFi access points (APs or routers) could boost efficacy in other data collection contexts, such as research studies. 
In addition, our findings have significant privacy implications, indicating that for practical purposes WiFi data should be considered location data.
As we argue in the following sections, this finding is not recognized in current practices of data collection and handling.




\section*{Results}

Ubiquitously available WiFi access points can be used as location beacons, identifying locations based on BSSID (basic service set identifier, uniquely identifying every router) broadcast by APs.
These locations are not intrinsically geographical, as the APs do not have geographical coordinates attached.
However, since the placement of APs tends to remain fixed, mapping an AP to a location where it was seen once is sufficient to associate all the subsequent scans from the user device with geographical coordinates.
See Supplementary Information for details on inferring the geographical locations of routers, as well as identifying (and discarding data from) mobile access points.

\begin{figure}[h!]
\begin{center}
\includegraphics[width=1\columnwidth]{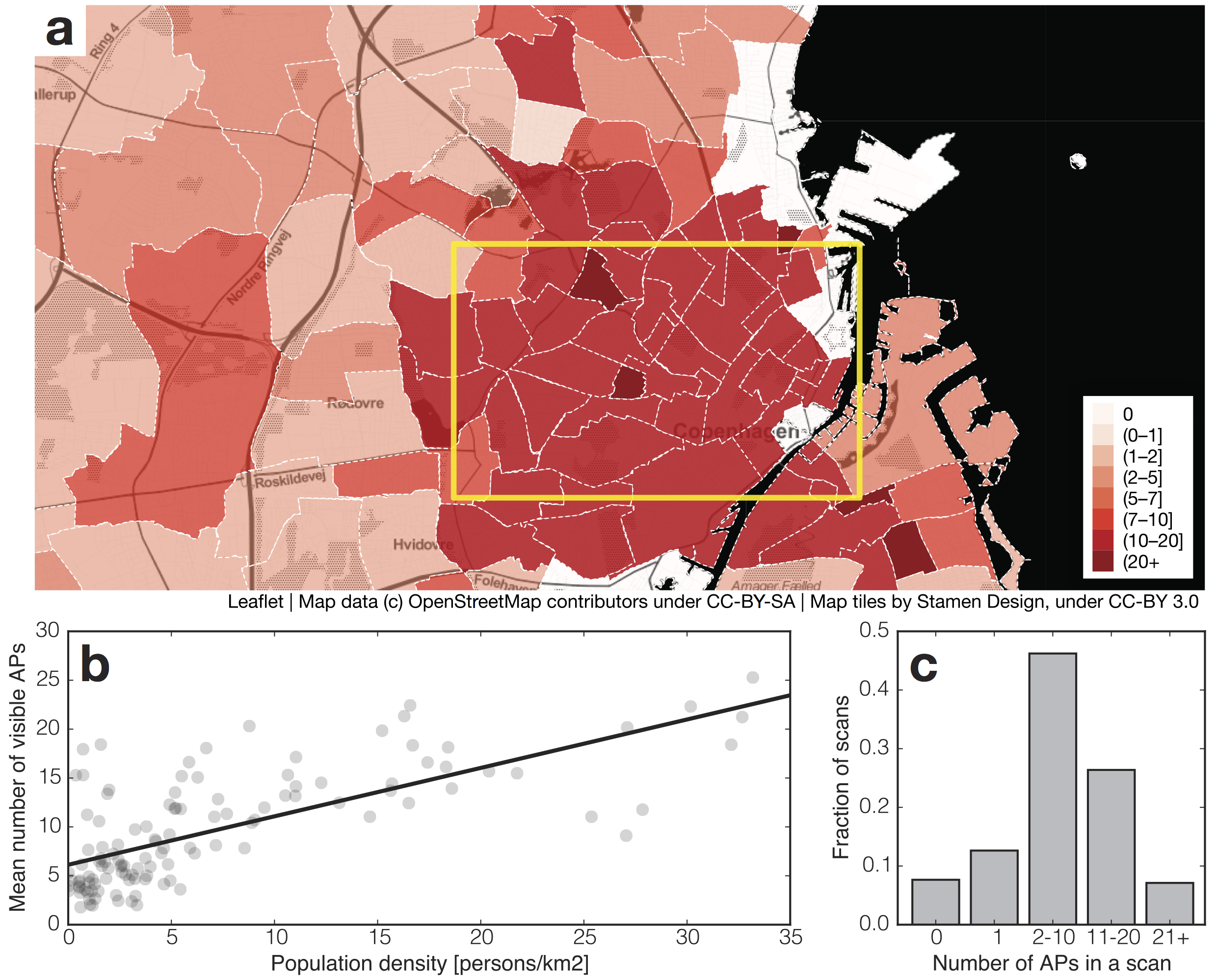}
\end{center}
\caption{\textbf{WiFi routers are located where people live.} 
\textbf{a:} Map of Greater Copenhagen Area, divided into parishes with color indicating average number of routers discovered per scan; rectangle overlay indicates the city center. 
\textbf{b: }The number of access points visible in each scan is correlated with the population density ($r^2=0.5$). 
Even in low population density areas there are several routers visible on average in each scan. 
Therefore, knowing the positions of only a subset of routers is enough for precise location sensing. 
\textbf{c: }Distribution of number of routers per scan. In our dataset 92\% of scans contain at least one router.}
\label{fig:parishes_map}
\end{figure}

WiFi networks are ubiquitous. In our population, 92\% of all WiFi scans detect at least one access point, and 33\% detect more than 10 APs, as shown in Figure~\ref{fig:parishes_map}c.
In densely-populated areas, an average of 25 APs are visible in every scan, with population density explaining 50\% of the variance of the number of APs, as shown in Figure~\ref{fig:parishes_map}b. 
WiFi scans containing at least one visible AP can be used for discovering the location of the user, with a typical spatial resolution on the order of tens of meters.

We investigate three approaches to using access points as location beacons, all of which enable WiFi-based location tracking even with limited resources: (1) recovering APs' locations from mobility traces collected during an initial training period (exploiting the long-term stability of human mobility), (2) recovering APs' locations from randomly sampled GPS updates (exploiting low entropy of human mobility, see Supplementary Information for distinction between stability and low entropy), and (3) using only the most frequently observed APs for which location can be feasibly obtained from external databases.
The task is to efficiently assign geographical coordinates (latitude and longitude) to particular APs, so they can be used as beacons for tracking user's location.
In the following sections, we refer to \emph{time coverage} as the fraction of ten-minute timebins, in which at least one router with a known location is observed. 

\subsection*{Stability of human mobility allows for efficient WiFi-based positioning.}
Human mobility has been shown to remain stable over long periods of time\csp\cite{gonzalez2008understanding}.
We find that participants in our study have stable routines, with locations visited in the first one, two, three, and four weeks of the study still visited frequently six months later.  
Learning the locations of routers seen during the first seven days (corresponding to ${\sim}$3.5\% of the observations, shown in Figure~\ref{fig:coverage}a, left panel) provides APs' locations throughout the rest of the experiment sufficient for recovering ${\sim}$55\% of users mobility until the Christmas break around days 75-90.
When the location of routers seen by each person is inferred using only this person's data (the personal-only WiFi database case, shown using an orange line in Figure~\ref{fig:coverage}), the information expires with time: there is a stable decrease in time coverage after Christmas break.
This decline is evident both when a week (Figure~\ref{fig:coverage}a, left panel) and four weeks (right panel) are used for training, with the time coverage dropping ${\sim}$18 percentage points between days 60 and 160.
The histograms above each plot show the distribution of time coverage in selected points in time (at 7, 80, 190 days respectively). 
The distribution for day 190 reveals that the expiry of the personal database validity is driven by individuals who significantly altered routines, with 40\% of participants spending only around 10\% of time in locations they have visited in the first week.
In contrast, when the inferred locations of routers are shared among people (the global database case, represented by a blue line) the information does not expire and shows no decreasing trend during the observation period.
This implies that rather than moving to entirely new locations, people begin to visit places that are new to them, but familiar to other participants. 
The histograms of time coverage distribution in both panels of Figure~\ref{fig:coverage}a reveal that the individuals are heterogeneous in their mobility.
The coverage in most cases is highly affected in the non-personal case (where the person does not collect their own location information, but data from others is used, marked using green in the figures), but 20\% of participants retain a coverage of above 80\% throughout the observation period, see Figure~\ref{fig:coverage}a, left panel.
People living and working close to each other (like students in a dormitory) share a major part of their mobility and thus location of the APs they encounter can be estimated using data collected by others.

The demonstrated stability of human mobility patterns over long periods has real-life privacy implications. 
Denying a mobile application access to location data, even after a short period, may not be enough to prevent it from tracking user's mobility, as long as its access to WiFi scans is retained.

\subsection*{Human mobility can be efficiently captured using infrequent location updates.}


 
Sampling location randomly across time (Figure~\ref{fig:coverage}b), rather than through the initial period (Figure~\ref{fig:coverage}a) provides a higher time coverage, which is retained throughout the observation.
With around one sample per day per person on average, the location can be inferred 80\% of the time in case of global lookup base and 70\% in personal case (see Figure~\ref{fig:coverage}c, at training fraction of 0.007).

The histograms in Figure~\ref{fig:coverage}b confirm that distribution of coverage in the non-personal case is bimodal within our population: 
mobility of some individuals can effectively be modeled using data from people around them, while patterns of others are so distinct they require using self-collected information. 
The single-mode distribution of coverage in the personal case and the fact that the distribution is unchanged between day 7 and day 190 show the lack of temporal decline when sampling happens throughout the observation period.

The GPS sensor on a mobile device constitutes a major battery drain when active\csp\cite{gps_consumption}, whereas the WiFi frequently scans for networks by default.
Our results show that GPS-based location sampling rate can be significantly reduced in order to save battery, while retaining high resolution location information through WiFi scanning.
Our analyses also point to another scenario where WiFi time series can result in leaks of personal information.
Infrequent location data can be obtained from a person's (often public) tweets, Facebook updates, or other social networking check-ins and then matched with their WiFi records to track their mobility.

\subsection*{Overall human mobility can be effectively captured by top WiFi access points.}

As previously suggested\csp\cite{song2010limits}, people's mobility has low entropy and thus a few most prevalent routers can work effectively as proxies for their location.
Figure~\ref{fig:coverage}d shows that inferring the location of just 20 top routers per person on average (which, given the median count of 22\,000 routers observed per person, corresponds to 0.1\% of all routers seen) translates to knowing the location of individuals 90\% of the time.
Since our population consists of students, who attend classes in different lecture halls in various buildings across the campus, we expect that the number of access points necessary to describe mobility of persons with a fixed work location can be even lower. There are persons in our study, for whom just four access points correspond to 90\% of time coverage (see Suppelementary~Figure~4 for details).

That the mobility of individuals in our sample overlaps is apparent in Figure~\ref{fig:coverage}d as the time coverage of three top routers in the personal case is the same as in the global coverage using the total of 80 routers (instead of 189 disjoint routers).

As a consequence, a third party with access to records of WiFi scans and no access to location data, can effectively determine the location of each individual 90\% of time by sending less than 20 queries to commercial services such as Google Geolocation API or Skyhook.

\subsection*{Single-user analysis.}
\begin{figure}[h!]
\begin{center}
\includegraphics[width=1\columnwidth]{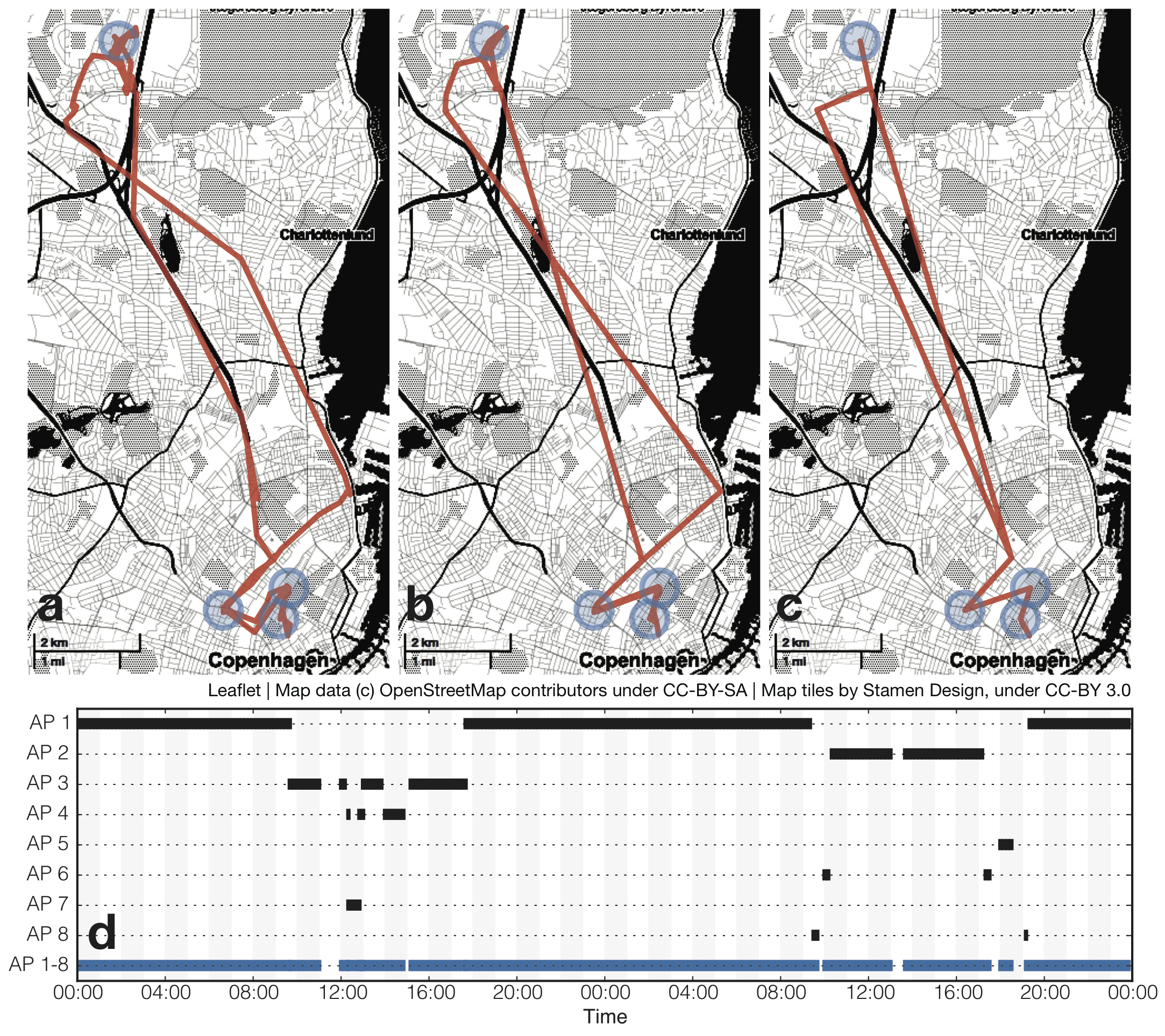}
\end{center}
\caption{\textbf{48 hours of location data of one of the authors, with the four visited locations visited marked in blue: home, two offices, and a food market.} Even though the author's phone has sensed 3\,822 unique routers in this period, only a few are enough to describe the location more than 90\% of time. {\bf a.} traces recorded with GPS; {\bf b.} traces reconstructed using all available data on WiFi routers locations - the transition traces are distorted, but all stop locations are visible and the location is known 97\% of the time. {\bf c.} with 8 top routers it is still possible to discover stop locations in which the author spent 95\% of the time. In this scenario transitions are lost. {\bf d.} timeseries showing when during 48 hours each of the top routers were seen. It can be assumed that AP 1 is home, as it's seen every night, while AP 2 and AP 3 are offices, as they are seen during working hours. The last row shows the combined 95\% of time coverage provided by the top 8 routers.}
\label{fig:subsampling}
\end{figure}

To illustrate the ubiquity of WiFi access points and how effectively they can be used to infer mobility patterns, we present a small example dataset containing measured and inferred location information of one of the authors, collected over two days.
During the 48 hours of observation, the researcher's phone was scanning for WiFi with a median period of 44 seconds, measuring on average 19.8 unique devices per scan, recording 3\,822 unique access points.
Only one scan during the 48 hours was empty, and one scan yielded 113 unique results.
Figure~\ref{fig:subsampling}a shows the corresponding GPS trace collected with a median sampling period of 5 minutes. 
When dividing the 48 hours of the test period into 10 minute bins, a raw GPS trace provides location estimation in 89\% of these bins. 
Four stop locations are marked with blue circles and include home, two offices, and a food market visited by the researcher.
Figure~\ref{fig:subsampling}b shows the estimation of this trace based on the inferred locations of WiFi routers, see Supplementary Information for detailed information on the location inference.
The four stop locations are clearly visible, but the transitions have lower temporal resolution and errors in location estimations.
This method provides location information in 97\% of temporal bins.
Using WiFi increases overall coverage, but might introduce errors in location estimation of routers which were only observed shortly, for example during transition periods.
Figure~\ref{fig:subsampling}c shows the estimation of this trace based on the locations of top 8 (0.2\%) WiFi routers. 
The four important locations have been correctly identified, but information on transitions is lost.
Information in 95\% of temporal bins is available.
Finally, Figure~\ref{fig:subsampling}d shows a graphical representation of how much time the researcher spends in any one of the top eight locations during the observation time.
Note that the first four locations account for an overwhelming fraction of the 48 hours.

Knowing the physical position of the top routers and having access to WiFi information reveals the location of the user for the majority of the timebins.
The details of trajectories become lost as we decrease the number of routers we use to estimate locations. 
With too few routers might not be possible to determine which of possible routes the subject chose or how long she took to travel through each segment of the trip.
On the other hand, the high temporal resolution of the scans allows for very precise discovery of arrival and departure times and of time spent in transit.
Such information has important implications for security and privacy, as it can be used to discover night-watch schedules, find times when the occupants are not home, or efficiently check work time of the employees.

\section*{Discussion}

Our world is becoming progressively more enclosed in infrastructures supporting communication, mobility, payments, or advertising.
Logs from mobile phone networks have originally been considered only for billing purposes and internal network optimization; today they constitute a global database of human mobility and communication networks\csp\cite{gonzalez2008understanding}.
Credit card records form high-resolution traces of our spending behaviors\csp\cite{krumme2013predictability}.
The omnipresent WiFi networks, intended primarily for communication, has now became a location tracking infrastructure, as described here.
The pattern is clear: every new cell tower, merchant with credit card terminal, every new private or municipal WiFi network offer benefits to the connected society, but, at the same time, create opportunities and perils of unexpected tracking.
Cities entirely covered by WiFi signal provide unprecedented connectivity to citizens and visitors alike; at the same time multiple parties have to incorporate this fact in their policies to limit privacy abuse of such infrastructure.
Understanding and quantifying the dynamics of privacy and utility of infrastructures is crucial for building connected and free society.

Since the creation of comprehensive databases containing geolocation for APs is primarily carried out by large companies\csp\cite{skyhook}, one might assume that WiFi based location tracking by `small players', such as research studies or mobile applications, is not feasible.
As we have shown above, however, APs can be very efficiently geolocated in a way that covers a large majority of individuals' mobility patterns.

In the results, we focused on outdoor positioning with spatial resolution corresponding to WiFi AP coverage: we assume that if at least one AP is discovered in a scan, we can assign the location of this AP to the user.
This is a deliberately simple model, described in detail in Supplementary Information,
but we consider the resulting spatial resolution sufficient for many aspects of research, such as studying human mobility patterns.
The spatial resolution of dozens of meters is higher than for example CDR data\csp\cite{gonzalez2008understanding}, which describes the location with the accuracy of hundreds of meters to a few kilometers.
Incorporating WiFi routers as location beacons can aid research by drastically increasing temporal resolution without additional cost in battery drain.

Students live in multiple dormitories on and outside of campus, take multiple routes commuting to the university, frequent different places in the city, travel across the country and beyond.
While the students spend most of their time within a few dozens of kilometers from their homes, they also make international and intercontinental trips (see Supplementary~Figures~2 and~3 for details).
Such long distance trips are not normally captured in studies based on telecom operator data.
Our population is densely-connected and in this respect it is biased, in the same sense as any population of people working in the same location.
We do simulate a scenario in which the individuals do not form a connected group by analyzing the results for personal-only database.
We expect the obtained results to generalize outside of our study.

Our findings connect to an ongoing debate about the privacy of personal data\csp\cite{strandburg2014privacy}.
Location data has been shown to be among the most sensitive categories of personal information\csp\cite{staiano2014money}.
Still, a record of WiFi scans is, in most contexts, not considered a location channel.
In the Android ecosystem, which constitutes 85\% of global smartphone market in Q2 2014\csp\cite{android}, the permission for applications to passively collect the results of WiFi scans is separate from the location permission; 
moreover, the \emph{Wi-Fi connection information (ACCESS\_WIFI\_STATE)} permission is not considered `dangerous' in the Android framework, whereas both high-accuracy and coarse location permissions are tagged as such\csp\cite{android_permissions}.
While it has been pointed out that Android WiFi permissions may allow for inference of sensitive personal information\csp\cite{achara2014wifileaks}, the effect has not been quantified through real-world data.
Here we have shown that inferring location with high temporal resolution can be efficiently achieved using only a small percentage of the WiFi APs seen by a device.
This makes it possible for any application to collect scanned access points, report them back, and inexpensively convert these access points into users' locations.
The impact is amplified by the fact that apps may passively obtain results of scans routinely performed by Android system every 15--60 seconds.
Such routine scans are even run when the user disables WiFi.
See Supplementary Information for additional analysis on data privacy in the Android ecosystem.

Developers whose applications declare both location and WiFi permissions are able to use WiFi information to boost the temporal resolution of any collected location information.
We have shown that even if the location permission is revoked by the user, or removed by the app developers, an initial collection of both GPS and WiFi data is sufficient to continue high-resolution tracking of the user mobility for subsequent months.
Many top applications in the Play Store request \emph{Wi-Fi connection information} but not explicit location permission.
Examples from the top charts include prominent apps with more than 100 million users each, such as Candy Crush Saga, Pandora, and Angry Birds, among others.
We are not suggesting that these or other applications collect WiFi data for location tracking.
These apps, however, do have a \emph{de~facto} capability to track location, effectively circumventing Android permission model and general public understanding. 

Due to uniqueness of location traces, users can be easily identified across multiple datasets\csp\cite{de2013unique}.
Our results indicate that any application can use WiFi permission to link users to other public and private identities, using data from Twitter or Facebook (based on geo-tagged tweets and posts), CDR data, geo-tagged payment transactions; in fact any geo-tagged data set.
Such cross-linking is another argument why WiFi scans should be considered a highly sensitive type of data.

In our dataset, 92\% of WiFi scans have at least one visible AP.
Even in the most challenging scenario, when there are no globally shared locations and each individual frequents different places, top 20 WiFi access points per person can be efficiently converted into geolocations (using Google APIs or crowd-sourced data) and used as a stable location channel.
These results should inform future thinking regarding the collection, use, and data security of WiFi scans. 
We recommend that WiFi records be treated as strictly as location data.

\section*{Methods}
\label{methods}

\subsection*{The dataset.}
Out of the 130+ participants of the study\csp\cite{10.1371/journal.pone.0095978}, we selected 63 for which at least 50\% of the expected data points are available.
The methods of collection, anonymization, and storage of data were approved by the Danish Data Protection Agency, and complies both with local and EU regulations. 
Written informed consent was obtained via electronic means, where all invited participants read and digitally signed the form with their university credentials.
The median period of WiFi scans for these users was 16 seconds, and the median period of GPS sampling was 10 minutes.
The data spans a period of 200 days from October 1st, 2012 to April 27th, 2013.

\subsection*{\emph{Known routers} and \emph{coverage}}
In the article we use a simple model of locating the WiFi routers. 
We consider an access point as \emph{known} if it occurred in a WiFi scan within one second of a GPS location estimation.
The shortcomings of this approach and possible remedies are described in more detail in Supplementary Information.

We define \emph{time coverage} as a fraction of ten-minute bins containing WiFi data in which at least one \emph{known router} was scanned.
For example, let us assume that the user has data in 100 out of 144 timebins during a day, and in 80 of these timebins there is a known router visible. Therefore, that user's coverage for that day is 80\%. The average time coverage for a day is the mean coverage of all users who had any WiFi information in that day.
This way our results are independent from missing data caused by imperfections in data collection system deployed in the study.

\begin{figure}[h!]
\begin{center}
\includegraphics[width=1\columnwidth]{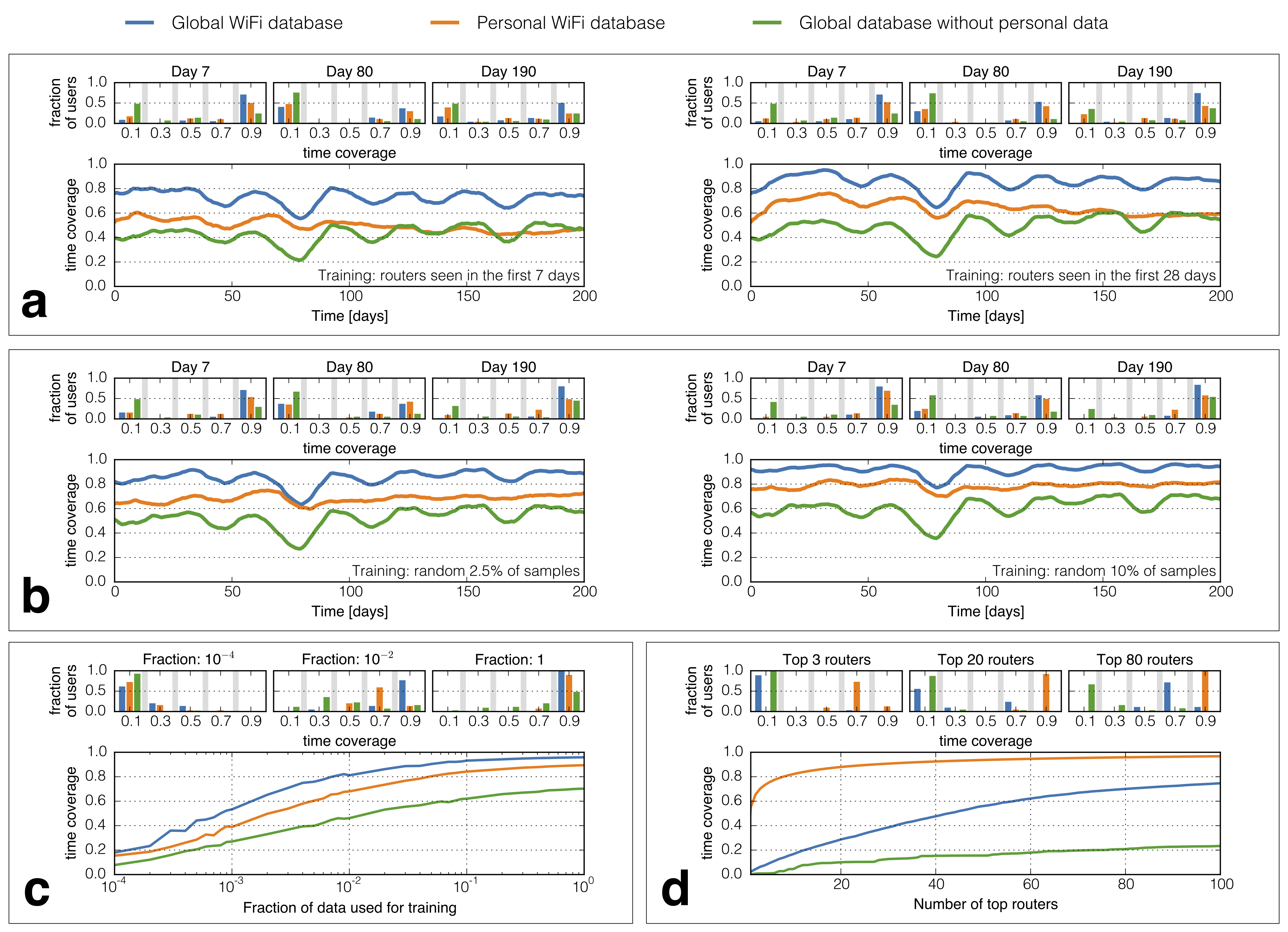}
\end{center}
\caption{\textbf{The time coverage provided by the routers with known position depends on who collects the corresponding location data and when it happens.} In each subplot the orange line describes the scenario where each individual collects data about themselves and does not share it with others; the blue line corresponds to a system in which the location of routers discovered by one person is made known to other users; the green line presents a situation where each individual can use the common pool of known routers but does not discover access points herself.
{\bf a. Stability of location.} 
Learning the location of APs seen during the first seven (left panel) or 28 (right panel) days, leads to performance gradually decreasing with time in the personal case (orange line). 
The histograms of time coverage distribution for day 190 show that this decline is driven by a growing number of people who spend only ${\sim}$10\% of time in the locations they visited in the beginning of the observation.
The global approach (blue line) does not show this tendency, which indicates that people rotate between common locations rather than moving to entirely new places.
{\bf b, c. Representativeness of randomly selected locations.} 
Random subsampling with an average period of 24 hours (less than 1\% of all available location estimations) is sufficient to find the most important locations in which people spend more than 80\% of their time; using an average period of 4 hours (2.5\% percent of all available location data) results in ${\sim}$85\% coverage. 
The personal database does not expire since the location is sampled throughout the experiment, not only in the beginning.
{\bf d. Limited number of important locations.} Although querying commercial services for WiFi geolocation is costly, knowing the location of only the 20 most prevalent routers per person in the dataset results in an average coverage of ${\sim}$90\%. Since people's mobility overlaps, there is a benefit of using a global database rather than treating all mobility disjointly.}
\label{fig:coverage}
\end{figure}

In Figure~\ref{fig:coverage} we present three different approaches to sampling, which we describe here in detail.
\textbf{Initial-period sampling.} 
As presented in Figure~\ref{fig:coverage}a, we learn the location of the routers sequentially.
With each GPS location estimation accompanied with a WiFi scan, we add the visible access points to the list of known routers.
The learning curve can be observed for the first seven days (Figure~\ref{fig:coverage}a, left panel) or the first 28 days (Figure~\ref{fig:coverage}a, right panel).
\textbf{Random subsampling.}
In the random subsampling scenario we select a set fraction of available GPS location estimations, each paired with a WiFi scan.
Each GPS estimation provides information on the position of all routers seen in the paired scan.
This scenario can be realized after the data collection is finished, as the location estimations are used to locate the WiFi scans which happened both before and after said estimations. 
The results are presented in Figure~\ref{fig:coverage}b.
\textbf{Top routers.}
We select the top routers in a greedy fashion after the data collection is finished.
We sort the routers in descending order by the number of user timebins they occur in.
We choose the top one router, and then we select the routers which provide the biggest increase in the number of user timbebins covered.
Due to high density of access points, each semantic place is described by presence of several routers, but location of only one of them has to be established to find the geographic position of the place.
In this sampling method we do not rely on our own GPS data --- top routers are found purely based on their occurrence in the WiFi scans, regardless of availability of GPS scans within the one second time delta.
The results of such sampling are presented in Figure~\ref{fig:coverage}e.

\subsection*{Data collection scenarios.}
Each subplot in Figure~\ref{fig:coverage} contains series coming from three different simulated collection scenarios.
In the \textbf{global} scenario, there is a pool of WiFi routers locations estimations coming from all users, and a router is considered known if at least one person has found its location. 
This scenario simulates the function of such services as for example mobile Google Maps. 
In the \textbf{personal} scenario each user can only use their own data, a router can be known to them only if they found its location themselves. 
It simulates collecting data in a disjoint society, where each person frequents different locations. 
Finally, in the \textbf{global with no personal data} scenario, each user can exploit estimations created by everybody else, but without contributing their own data.

{\footnotesize 
\noindent \textbf{Acknowledgements.} We thank Yves-Alexandre de Montjoye and Andrea Cuttone for useful discussions. \\
\textbf{Funding.} The research project is supported by the Villum Foudation, as well as the University of Copenhagen through Programmes of Excellence for Interdisciplinary Research. \\
\textbf{Contributions.} Designed the research - SL, RG, PS, AS; Analysed the data - PS, RG; wrote and reviewed the manuscript - PS, AS, RG, SL. \\
\textbf{Additional Information.}  The authors declare no competing financial interests.
}

\FloatBarrier
\clearpage
\setcounter{figure}{0} 

\renewcommand{\figurename}{Supplementary Figure}

\section*{Supplementary Information}
\subsection*{Inferring location of routers.}
In the article we use a deliberately simplistic model of locating the WiFi routers. 
We assume that if we find a WiFi scan and a GPS location estimation which happened within a one second time difference we can assume that all routers visible in the scan are at the geographical location indicated by the GPS reading.
Due to effective outdoor range of WiFi routers of approximately 100~meters, this assumption introduces an obvious limitation of accuracy of location inference. 
Moreover, there are a number of mobile access points such as routers installed in public transportation or smartphones with hotspot capabilities. 
Such devices cannot effectively be used as location beacons and will introduce noise into location estimations unless identified and discarded.
We propose and test the following method.
For each GPS location estimation with timestamp $t_{GPS}$ we find WiFi scans performed by the same device at $t_{WiFi}$ so that $t_{GPS}-1s \leq t_{WiFi} \leq t_{GPS}+1s$ and select the one, for which $|t_{GPS}-t_{WiFi}|$ is the smallest.
We then add the location estimation and its timestamp to the list of locations where each of the available WiFi access points was seen.
For each device, we fit a density-based spatial clustering of applications with noise (DBSCAN) model\csp\cite{dbscan} specifying 100 meters as the maximum distance parameter $\varepsilon$.
If there are no clusters found, or the found clusters contain less than 95\% of all locations associated with the said router we assume the router is mobile and to be discarded from further analysis.
If only one cluster is detected and it contains at least 95\% of all points, we assume the geometric median of these points is the physical location of the router.
If there are more clusters found and they contain at least 95\% of all points, we verify if these clusters are disjoint in time:
if the timestamps of sightings do not overlap between those clusters, we assume the device is a static access point which has been moved to a different place during the experiment. 
Otherwise, we classify the access point as a mobile device and do not use it as a location proxy.

\begin{figure}[h]
\begin{center}
\includegraphics[width=1\columnwidth]{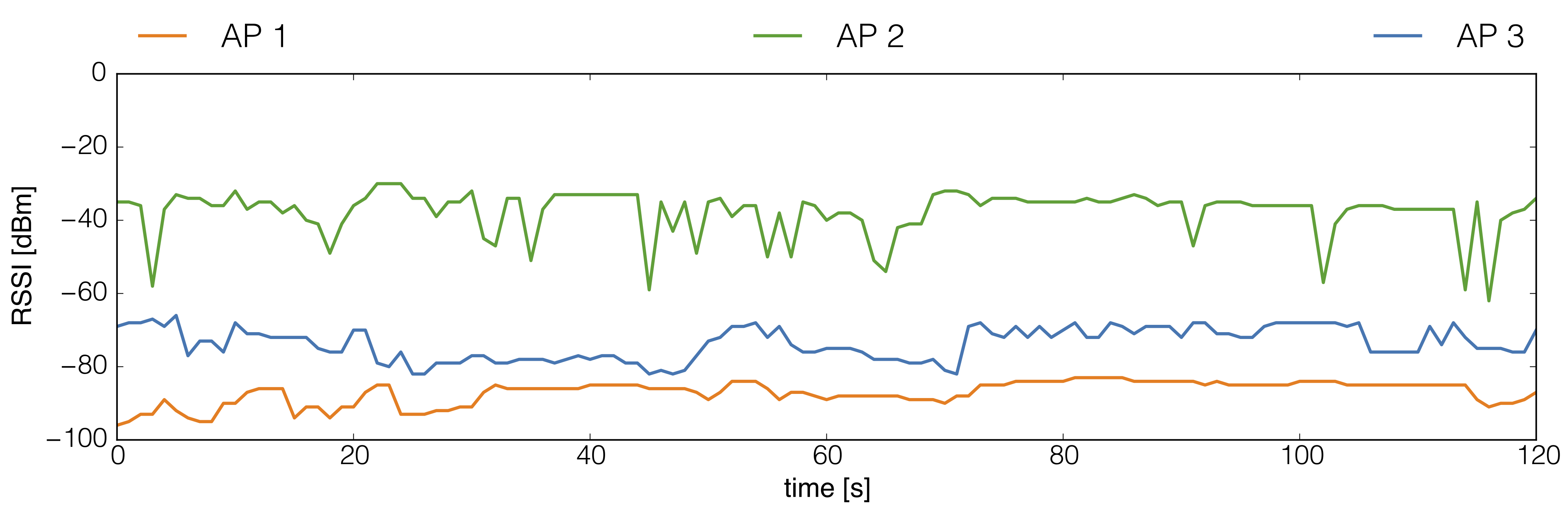}
\end{center}
\caption{Received signal strength can vary greatly even if the smartphone and the access points do not move.}
\label{fig:wifi_rssi}
\end{figure}

In the proposed method we assume accuracy of tens of meters is satisfactory, and hence do not find a need to exploit the received signal strength information\csp\cite{liu2007survey}. 
Arguably, with the sparse data that we operate on, employing received signal strength could lead to more confusion, as it can vary greatly for one location, depending on the position of the measuring smartphone, and presence of humans and other objects obstructing the signal. 
Supplementary Figure~\ref{fig:wifi_rssi} shows timeseries of signal strengths received by a non-moving smartphone, which vary as much as 10\,dB, which corresponds to drastic differences in estimated distance to the source, as in free-space propagation model extending the distance $\sqrt{10} \approx 3.16$ times corresponds to 10\,dB loss in received signal strength. 

In the 200 days of observations, the participants have scanned 487\,216 unique routers, out of which 64\,983 were scanned within a second of a GPS estimation.
As many as 57\,912 were only seen less than five times which we assumed to be the minimum number of sightings to be considered a cluster, which left only 7\,071 routers for further investigation.
In 1\,760 cases there were no clusters found, or there was more than 5\% noise.
In 5\,267 cases there was only one cluster and less than 5\% of noise.
Out of 21 cases there were multiple clusters and less than 5\% of noise, 9 revealed no time overlap between clusters.
We verified our heuristic of determining which routers are mobile by classifying routers which are very likely mobile, as their networks are called AndroidAP (default SSID for a hotspot on Android smarphones), iPhone (default SSID for iPhones), Bedrebustur or Commutenet (names of networks on buses and trains in Copenhagen).
Out of 340 such devices 323, or 95\%, were identified as mobile, and 17 as fixed-location devices.

All in all, out of 487\,216 unique APs we believe we managed to estimate the location of 5\,276, we identified 1\,771 as mobile, and did not have enough data to investigate 480\,169.
Even though we only know the location of approximately 1\% of all sensed routers, this knowledge is enough to estimate the location of users in 87\% ten-minute timebins in the dataset.

\subsection*{Long term stability and low entropy of human mobility.}
Long-term stability in the context of human mobility means that individuals keep returning to the same locations over long time periods.
Arguably, most people do not often move, change the work place, or find an entirely new set friends to visit.
We use entropy in Shanon's definition, as presented in equation~(\ref{eq:entropy})
\begin{equation}
\label{eq:entropy}
H(X) = -\sum_{i}{P(x_i)\log{P(x_i)}},
\end{equation}
where $X$ is the set of all possible locations, and $P(x_i)$ is the probability of a person being at location $i$.
Therefore, the bigger the fraction of time a person spends in their top few places, the lower the entropy value of that person's mobility.
In this sense, long-term stability is necessary for the low entropy, and both contribute to the predictability of human mobility.

\subsection*{Mobility of the studied population.}
This article focuses on a population of students at a university. To show that their mobility is not constrained to the campus only, we present summary statistics about their mobility. Displacements in our dataset can be as big as 10\,000 km. Given such extreme statistics, the radius of gyration, while commonly used in literature to describe mobility on smaller scales~\cite{gonzalez2008understanding}, is not a suitable measure here. Instead, in Supplementary Figure~\ref{fig:world_map} we show a qualitative overview in form of a heatmap of observed locations, as well as a distribution of time spent as a function of distance from home. For simplicity, we define the home location for each student as the location of the most prevalent access point in their data. We then calculate the median distance from home for each hour of the observation using their location data. For a more detailed view, we present the distribution for 48 randomly chosen students in Supplementary Figure~\ref{fig:histograms}.

\begin{figure}[h]
\begin{center}
\hspace{29px}\includegraphics[width=.875\columnwidth]{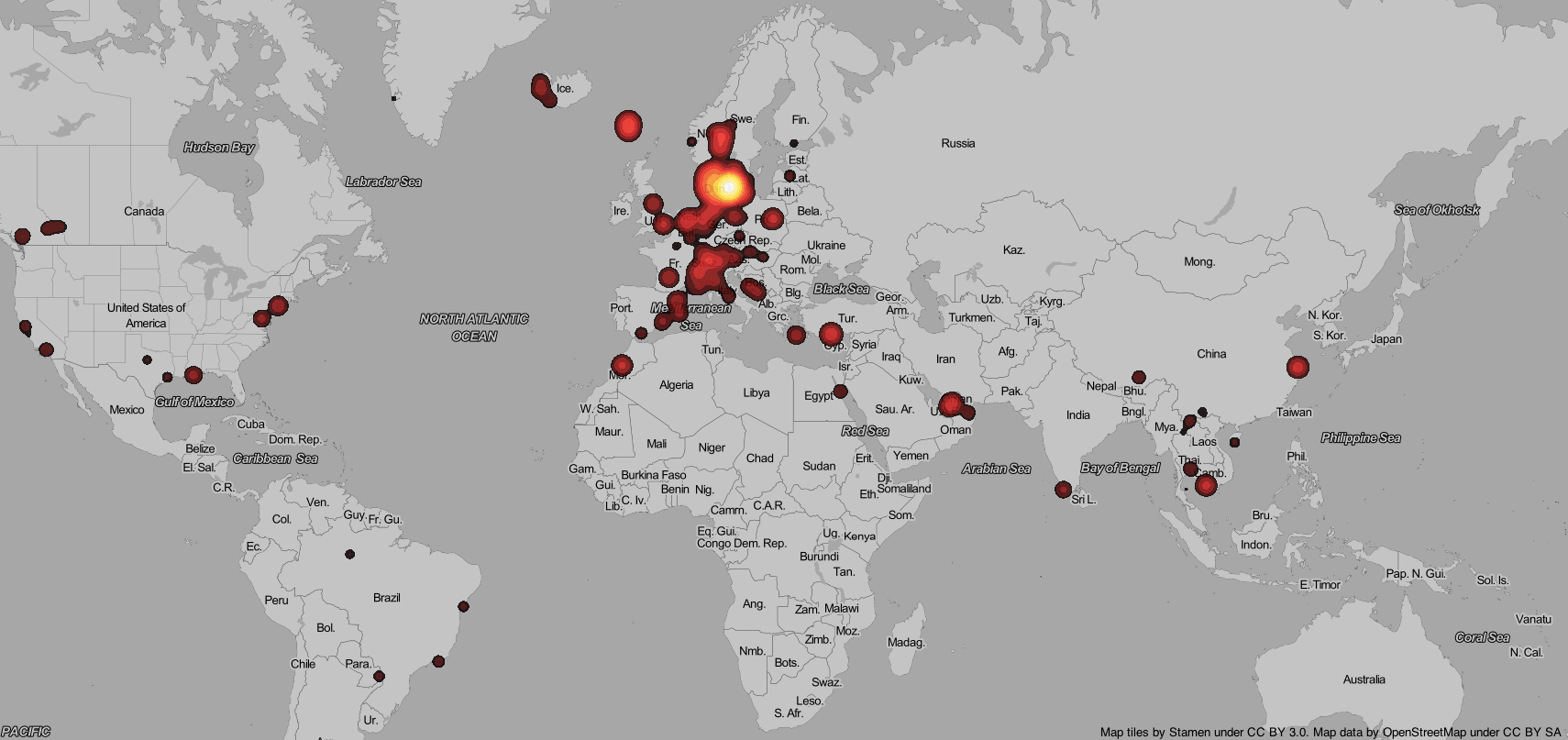}\\
\includegraphics[width=1\columnwidth]{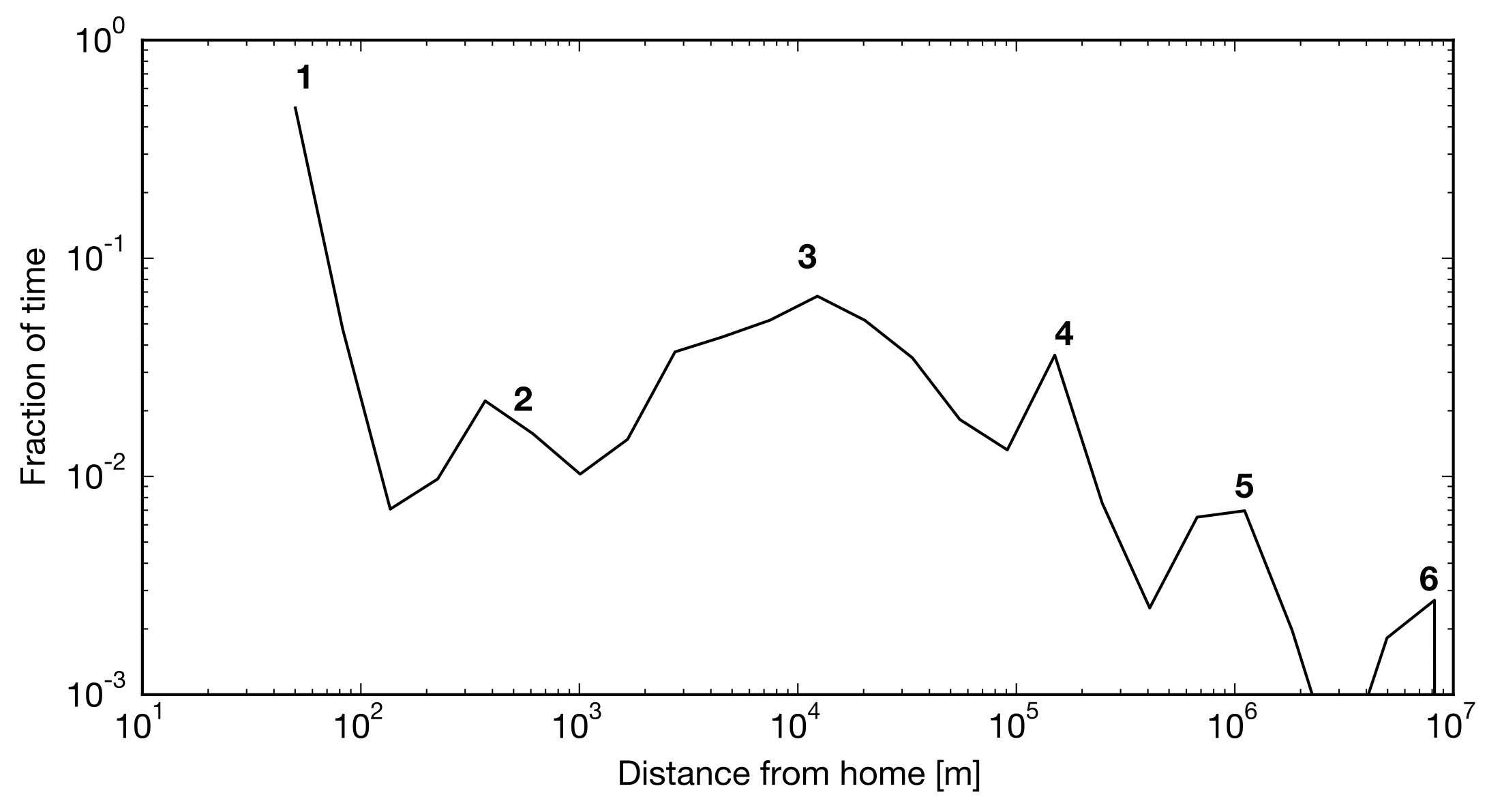}
\end{center}
\caption{The article focuses on a population of students at a single university, but they are not constrained to the campus only. Our data captures human mobility at different scales: the participants spend most of their time at home (1), but they travel around the neighborhood (2), the city (3), to different cities in Denmark (4), different cities in Europe (5), and finally, other continents (6).}
\label{fig:world_map}
\end{figure}

\begin{figure}[h]
\begin{center}
\includegraphics[width=\columnwidth]{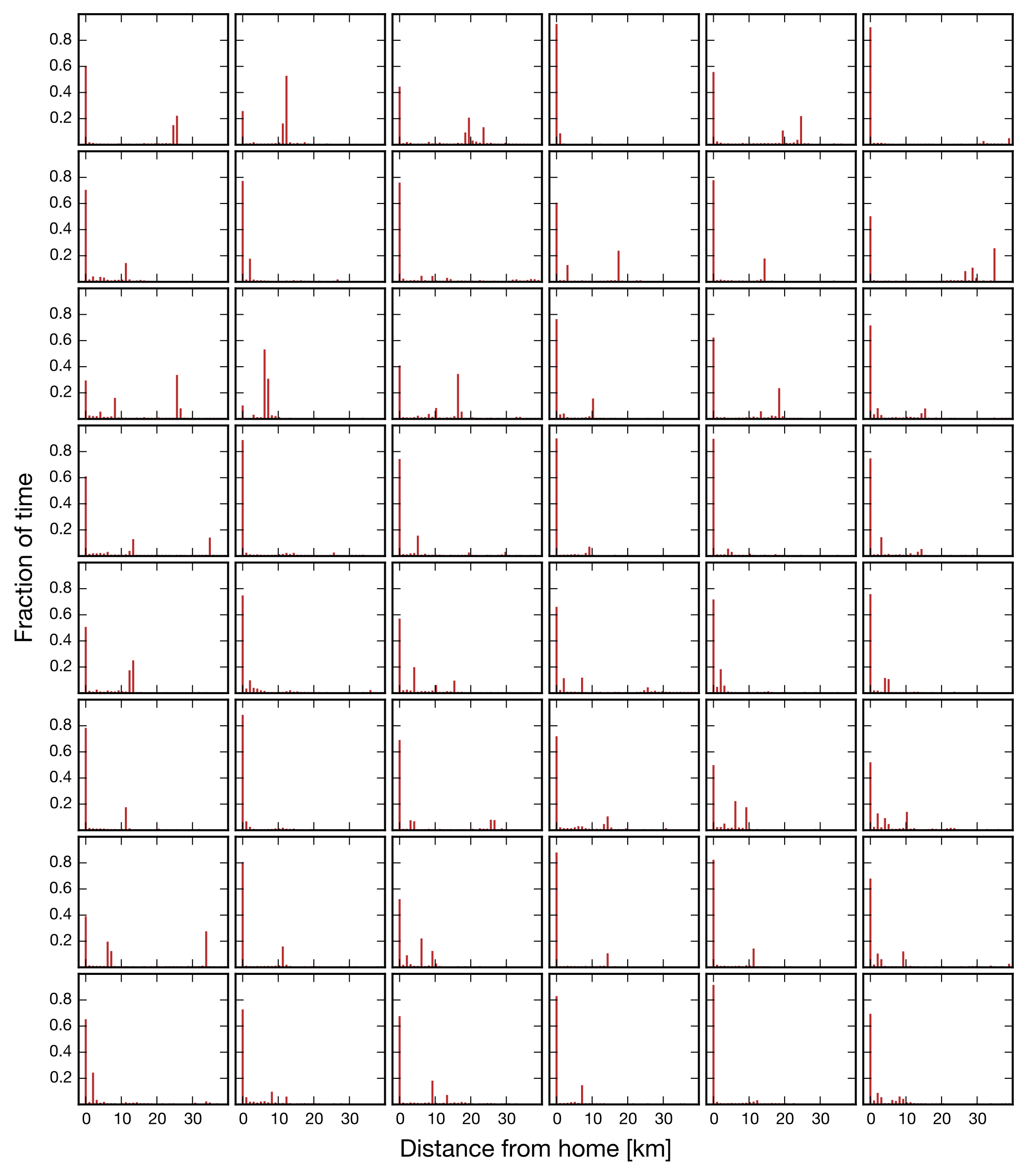}
\end{center}
\caption{Distribution of time spent at different distances from the inferred home location, presented for randomly selected 48 participants. In most cases, we see the home location as the most prevalent, and probably a "work" location as the next peak in the distribution.}
\label{fig:histograms}
\end{figure}

\subsection*{Time coverage of top routers.}
\begin{figure}[h]
\begin{center}
\includegraphics[width=1\columnwidth]{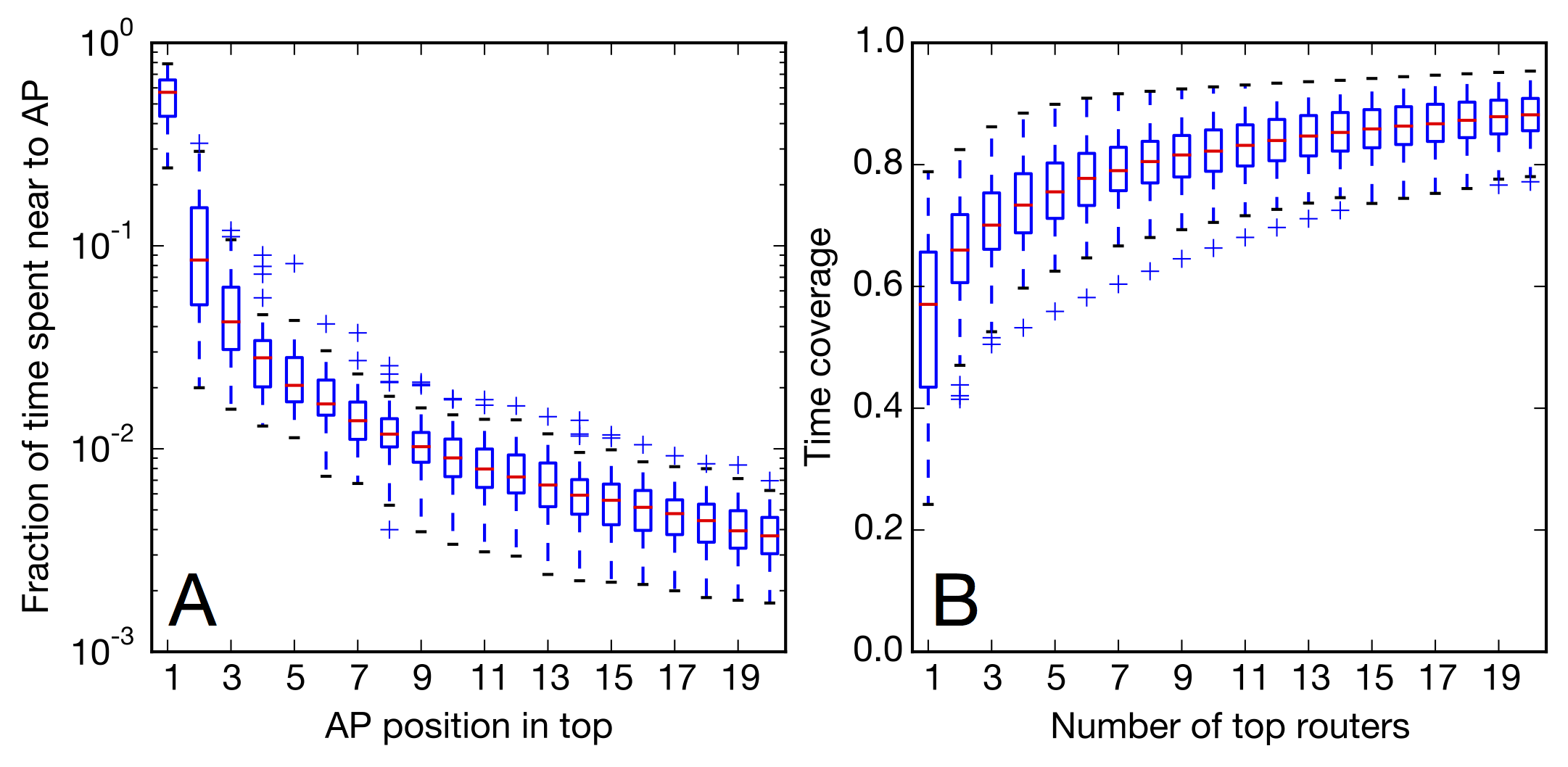}
\end{center}
\caption{A more detailed view of time coverage provided by top routers found through the greedy algorithm. A: there is a clear main location for a majority of participants, we therefore assume this to be the home location. B: even though 20 routers are needed on average to capture 90\% of mobility, there are participants for whom just four routers suffice.}
\label{fig:top_locations}
\end{figure}

In this section we present a more detailed view on time coverage of top routers selected separately for each person. Supplementary Figure~\ref{fig:top_locations}A shows the fraction of time which participants spent near to one of their top 20 routers. It is worth noting, that while home location is immediately apparent, there seems to be no definite "work" location in our population. This can be attributed to the fact that the participants of the observation are students who attend classes in different buildings and lecture halls and do not have an equivalent of an office. Supplementary Figure~\ref{fig:top_locations}B is an enriched version of Figure 2d from the main text of the article. It shows that even though 20 routers are needed on average to capture 90\% of mobility, there are participants for whom just four routers suffice.

\subsection*{Android Permissions.}

The scope of Android permission \emph{ACCESS\_WIFI\_STATE} is described in the developer documentation as ``allows applications to access information about Wi-Fi networks''\csp\cite{android_permission_1}.
This permission provides the requesting application with a list of all visible access points along with their MAC identifiers after each scan ordered by any application on the phone (via broadcast mechanism).
Moreover, with this permission the applications can start in the background when the first WiFi scan results appear after the phone boots: the app's BroadcastReceiver is called and the data can be collected without explicit \emph{RECEIVE\_BOOT\_COMPLETED} permission.
Requesting a WiFi scan requires the \emph{CHANGE\_WIFI\_STATE} permission, marked as dangerous, but in most cases it is not necessary to request it:
the Android OS by default performs WiFi scans in the intervals of tens of seconds, even when the WiFi is turned off; the setting to \emph{disable} background scanning when WiFi is off is buried in the advanced settings. 

Application developers often use \emph{ACCESS\_WIFI\_STATE} to obtain information whether the device is connected to the Internet via mobile or WiFi network.
This information is useful, for example, to perform larger downloads only when the user in connected to a WiFi network and thus avoid using mobile data.
This is an unnecessarily broad permission to use for this purpose, as the same information can be obtained with \emph{ACCESS\_NETWORK\_STATE}, which provides all the necessary information without giving access to personal data of WiFi scans:

{\footnotesize   
\begin{lstlisting}
ConnectivityManager cManager = 
	(ConnectivityManager) getSystemService(Context.CONNECTIVITY_SERVICE);
NetworkInfo mWifi = cManager.getNetworkInfo(ConnectivityManager.TYPE_WIFI);
if (mWifi.isConnected()) {  } //wifi is connected
\end{lstlisting}
}

Since the \emph{ACCESS\_WIFI\_STATE} together with \emph{INTERNET} permission (for uploading the results) are effectively sufficient for high-resolution location tracking, we suggest the developers transition to using the correct permissions and APIs for determining connectivity and that accessing the result of WiFi scan requires at least the \emph{ACCESS\_COARSE\_LOCATION} permission.

\renewcommand{\figurename}{Supplementary Figure} 
 \setcounter{figure}{0}
\clearpage{}\end{document}